\documentclass[aps,prd,twocolumn]{revtex4}
\usepackage{amssymb}
\usepackage{amsfonts}
\usepackage{amsmath}
\usepackage{graphicx}
\usepackage{color}
\usepackage[dvips]{epsfig}
\usepackage[dvips]{graphicx}
\usepackage{float}

\begin{document}

\title{Nontopological self-dual Maxwell-Higgs vortices}
\author{D. Bazeia$^{1}$, R. Casana$^2$, M. M. Ferreira Jr.$^2$, E. da Hora$^{2,3}$}

\affiliation{
  $^{1}$ Departamento de F\'{\i}sica, Universidade Federal da Para\'{\i}ba,
58051-970, Jo\~{a}o Pessoa, Para\'{\i}ba, Brazil.\\
  $^2$ Departamento de F\'{\i}sica, Universidade Federal do Maranh\~{a}o,
65085-580, S\~{a}o Lu\'{\i}s, Maranh\~{a}o, Brazil.\\
 $^3$Coordenadoria Interdisciplinar de Ci\^{e}ncia e Tecnologia,
Universidade Federal do Maranh\~{a}o, {65080-805}, S\~{a}o Lu\'{\i}s, Maranh\~{a}o, Brazil.}

\begin{abstract}
We study the existence of self-dual nontopological vortices in generalized
Maxwell-Higgs models recently introduced in Ref.~\cite{gv}. Our
investigation is explicitly illustrated by choosing a sixth-order
self-interaction potential, which is the simplest one allowing the existence
of nontopological structures. We specify some Maxwell-Higgs models yielding
BPS nontopological vortices having energy proportional to the magnetic flux,
$\Phi _{B}$, and whose profiles are numerically achieved. Particularly, we
investigate the way the new solutions approach the boundary values, from
which we verify their nontopological behavior. Finally, we depict the
profiles numerically found, highlighting the main features they present.
\end{abstract}

\maketitle

\section{Introduction}

\label{Intro}

This work deals with the presence of nontopological vortices in generalized
Maxwell-Higgs models recently introduced in Ref.~\cite{gv}, whose dynamics
is controlled by two positive functions depending on the Higgs field only.
Before focusing on this, however, it seems interesting to first comment on
topological solutions. As one knows, topologically nontrivial
configurations, generically named topological defects, are frequently
described as static solutions to some nonlinear classical field models
allowing for the spontaneous symmetry breaking mechanism \cite{n5}. In this
sense, such structures have been studied intensively all over the years,
their consequences being applied to many areas of interest, specially in
Cosmology, since they are known to be formed during symmetry breaking phase
transitions \cite{VLK}.

The interesting point to be raised is that, according to the usual approach,
these solutions can be found by solving a given set of first-order
differential equations (instead of the second-order Euler-Lagrange ones)
\cite{n4}. The main advantage of following such a prescription is that it
also reveals that the resulting solutions are stable against decaying into
the respective mesons (i.e. they saturate a lower bound for the energy of
the model), with their energy depending only on the physical fields boundary
conditions (i.e. the topology). The simplest solution\ is the static kink
coming from a (1+1)-dimensional model described by one real scalar field
only \cite{n0}. In addition, there are also vortices and magnetic monopoles:
the first ones are born in (1+2)-dimensional gauged Abelian-Higgs theory
\cite{n1}, while the second ones stand for the static solutions\ appearing
in (1+3)-dimensional gauged non-Abelian-Higgs models \cite{n3}.

Recently, many authors have also investigated topological configurations
coming from generalized models possessing nonusual dynamics. In these works
interesting new properties have been discovered, including variations on the
shape of the resulting solitons, see Refs. \cite{o2,Adam}. In addition, some
of these models have been verified to support first-order solutions and a
lower bound for their energy, see Ref. \cite{o1}. Here, it is important to
point out that the main motivation regarding noncanonical dynamics comes\ in
a rather natural way\ from the string theory, and that interesting
applications of such an idea have been found in connection with the
accelerated inflationary phase of the universe \cite{n8}, tachyon matter
\cite{tm}, dark matter \cite{dm}, and others \cite{o}.

In field theory models, finite energy solutions require that the
self-interacting potential goes to zero as the field solutions approach
their asymptotic profiles. Stable nontopological solutions satisfy
asymptotical boundary conditions that imply finite energy and null
topological charge (becoming stable due to conservation of Noether charges).
Nontopological defects with vorticity $n>0$ have null profiles at origin and
far from it, requiring a $\phi ^{6}$ potential, as it occurs in the
Chern-Simons-Higgs model \cite{cshv} and in some extended electrodynamics
\cite{Ghosh}. Nontopological solutions were initially studied in models
constituted of one and two scalar fields, with applications in QCD \cite%
{Birse,Lee}, see also Ref. \cite{Wilet}. Some investigations about the
conditions under which modified Lagrangians, composed of generalized kinetic
terms, yield nontopological soliton solutions were undertaken in Ref. \cite%
{Diaz}.

The impossibility to attain nontopological solutions in a Maxwell-Higgs
model is associated to the usual $\phi ^{4}$\ self-interacting potential,
which provides BPS solutions. A generalized Maxwell-Higgs model with
dielectric functions depending on the scalar field, was recently proposed
\cite{gv}, and may be used to achieve nontopological solutions. In the
present work, we investigate new aspects of the Maxwell-Higgs model of Ref.
\cite{gv}, showing that it also supports nontopological configurations
whenever supplemented with self-interacting potentials that engender
symmetric minimum.

In order to present our results, this work is organized as follows. In Sec.
II, we review the way the adopted nonusual Abelian-Higgs model engenders
self-duality, embracing the first-order equations and the lower bound for
the overall energy (defined in terms of an auxiliary function conveniently
introduced).\ In Sec. III, we define the scenarios we consider by choosing
the Higgs potential suitably. In addition, we characterize our solutions by
investigating the way they approximate the boundary values, which the
physical fields are supposed to reach (near the origin and asymptotically).
We also show that, in the appropriate limit, these models admit the very
same analytical solutions and related conclusions. We go further by
presenting numerical profiles, which explicitly describe the general
properties of the solutions. We still point out the existence of models
engendering\ entirely analytical nontopological solutions (i.e. not
approximate ones). Finally, in Section IV, we present our concluding remarks
and perspectives regarding future investigations.


\section{\textbf{The model}}

\label{general}

We begin by reviewing the first-order framework firstly introduced in Ref.
\cite{gv} for attaining topological solutions, whose starting-point is the $%
(1+2)$-dimensional generalized Maxwell-Higgs Lagrangian density,
\begin{equation}
\mathcal{L}=-\frac{G\left( \left\vert \phi \right\vert \right) }{4}F_{\mu
\nu }F^{\mu \nu }+w\left( \left\vert \phi \right\vert \right) \left\vert
D_{\mu }\phi \right\vert ^{2}-U\left( \left\vert \phi \right\vert \right)
 , \label{1}
\end{equation}%
where $F_{\mu \nu }=\partial _{\mu }A_{\nu }-\partial _{\nu }A_{\mu }$ is
the electromagnetic field strength tensor, $D_{\mu }\phi =\partial _{\mu
}\phi +ieA_{\mu }\phi $ is the minimal covariant derivative. Here, we use
standard conventions, including the plus-minus signature for the Minkowski
metric. The positive functions, $G\left( \left\vert \phi \right\vert \right)
$ and $w\left( \left\vert \phi \right\vert \right) $, generalize the
dynamics of the usual Maxwell-Higgs model. It is easy to show that the Gauss
law of this model is saturated by the temporal gauge, $A_{0}=0$, so it
describes only magnetic configurations, as the usual case.

Our aim is looking for radially symmetric configurations coming from (\ref{1}%
). For such purpose we implement the standard stationary vortex Ansatz,%
\begin{equation}
\phi \left( r,\theta \right) =\upsilon g\left( r\right) e^{in\theta } ,
\;\mathbf{A}\left( r,\theta \right) =-\frac{\widehat{\theta }}{er}\left(
a\left( r\right) -n\right)  , \label{2b}
\end{equation}%
where $n=\pm 1,\pm 2,\pm 3...$ stands for the winding number of the
topological configuration.

We focus our attention on the self-dual configurations of Lagrangian (\ref{1}%
), governed by two coupled first-order equations that minimize the total
energy of the model. In order to obtain such equations, we follow the usual
Bogomol'nyi-Prasad-Sommerfield (BPS) formalism by writing the energy density
of this model in the radially symmetric\ Ansatz,
\begin{equation}
\varepsilon =\frac{G}{2e^{2}}\left( \frac{1}{r}\frac{da}{dr}\right)
^{2}+\upsilon ^{2}w\left( \left( \frac{dg}{dr}\right) ^{2}+\frac{a^{2}g^{2}}{%
r^{2}}\right) +U , \label{5}
\end{equation}%
where $U=U\left( g\right) $. In the present scenario, self-duality only
holds when $G\left( \left\vert \phi \right\vert \right) $, $w\left(
\left\vert \phi \right\vert \right) $ and $U\left( \left\vert \phi
\right\vert \right) $ are constrained to each other by%
\begin{equation}
\frac{d}{dg}\sqrt{GU}=-\sqrt{2}e\upsilon ^{2}wg.  \label{6}
\end{equation}%
Under such a constraint, the energy density Eq. (\ref{5}) can be rewritten
as,
\begin{eqnarray}
\varepsilon  &=&\frac{G}{2}\left( \frac{1}{er}\frac{da}{dr}\pm \sqrt{\frac{2U%
}{G}}\right) ^{2}+\upsilon ^{2}w\left( \frac{dg}{dr}\mp \frac{ag}{r}\right)
^{2}  \nonumber \\
&&\mp \frac{1}{er}\frac{d}{dr}\left( a\sqrt{2GU}\right)  , \label{7}
\end{eqnarray}%
from which one concludes that the corresponding total energy saturates its
lower bound when $g\left( r\right) $ and $a\left( r\right) $ satisfy%
\begin{eqnarray}
\frac{dg}{dr} &=&\pm \frac{ag}{r} , \label{8} \\
\frac{1}{er}\frac{da}{dr} &=&\mp \sqrt{\frac{2U}{G}}.  \label{9}
\end{eqnarray}%
These are the BPS or self-dual first order equations of the generalized
model. Saturating the first-order equations, the energy density is shown to
be
\begin{equation}
\varepsilon _{bps}=\mp \frac{1}{er}\frac{dH}{dr} , \label{11}
\end{equation}%
where we have defined the auxiliary function
\begin{equation}
H\left( r\right) =a\sqrt{2GU}.  \label{Hf1}
\end{equation}%
By using BPS\ equations, we also can write the correspondent energy density
as%
\begin{equation}
\varepsilon _{bps}=2U+2\upsilon ^{2}w\left( \frac{ag}{r}\right) ^{2} ,\label{BPS_def_pos}
\end{equation}%
which will be positive definite for $w(g)>0$. In summary, given $G$, $w$\
and $U$\ constrained by relation (\ref{6}), the first-order equations (\ref%
{8}) and (\ref{9}) give rise to self-dual solutions for both $g(r)$\ and $%
a(r)$\ which satisfying the boundary conditions,
\begin{eqnarray}
g\left( 0\right)  &=&0 ,\;~g\left( \infty \right) =1 ,\\[0.2cm]
a\left( 0\right)  &=&n ,\;~a\left( \infty \right) =0 ,\end{eqnarray}%
engendering topological configurations. These solutions possess finite
energy obtained by integrating Eq. (\ref{11}) over the plane, i.e.%
\begin{equation}
E_{bps}=2\pi \int r\varepsilon _{bps}dr=\pm 2\pi \left[ H\left( 0\right)
-H\left( \infty \right) \right]  , \label{11a}
\end{equation}%
with $H\left( 0\right) $ and $H\left( \infty \right) =0$ being the values of
$H(r)$ at the boundaries. In this model, suitable choices of functions $G$\
and $w$\ allowed to achieve finite energy BPS solutions associated with
potentials $U$\ different from the usual $\phi ^{4}$\ one \cite{gv}, with
the energy resulting proportional to the magnetic flux. So far, no study
about the possibility of attaining BPS nontopological solutions in this
Maxwell-Higgs framework was developed.


\section{Nontopological vortices}

\label{general copy(1)}

In this Section, we show that the generalized Maxwell-Higgs model of Ref.
\cite{gv} may also support nontopological radially symmetric solitons,
assuming some specific choices for the generalized functions $G$\ and $w$,
and that the first-order equations (\ref{8}) and (\ref{9}) are solved
fulfilling suitable boundary conditions. We can show that the energy of
these BPS nontopological magnetic vortices is proportional to the magnetic
flux $\Phi _{B}$,\ not necessarily quantized in this case. From now on and
for simplicity, we choose $e=\upsilon =1$, and consider only the case $n>0$,
corresponding to the upper signs in Eqs. (\ref{8}), (\ref{9}), (\ref{11}), (%
\ref{11a}).

In order to obtain nontopological configurations, the profile functions $%
g\left( r\right) $ and $a\left( r\right) $ must obey the following boundary
conditions\ for $n\neq 0$,%
\begin{eqnarray}
g\left( 0\right)  &=&0 ,\;~g\left( \infty \right) =0 , \label{3}
\\[0.2cm]
a\left( 0\right)  &=&n ,\;~a\left( \infty \right) =-\alpha _{n} ,\label{4}
\end{eqnarray}%
where $\alpha _{n}$ \ stands for a positive real number calculated
numerically. Under such boundary conditions, the magnetic flux $\Phi _{B}$
is immediately obtained
\begin{equation}
\Phi _{B}=2\pi \int rB\left( r\right) dr=2\pi \left( n+\alpha _{n}\right)
 , \label{mf}
\end{equation}%
and is not necessarily quantized (i.e. $\alpha _{n}$ is not an integer).

Notwithstanding the magnetic flux be given by (\ref{mf}), it is still
possible to define classical field models for which the magnetic flux
defines the lower bound for the BPS energy (\ref{11a}). Note that this
occurs only when the function $H$\ satisfies boundary conditions
proportional to the ones of Eq. (\ref{4}), which requires a particular
choice for the functions $G$\ and $w$.

It is well-known that nontopological solutions arise in field theories,
which possess at least one symmetric vacuum, and asymptotical conditions
compatible with finite energy. In canonical field models, the simplest
Higgs-potential consistent with this requirement is the usual sixth-order
one,
\begin{equation}
U\left( g\right) =\frac{\lambda ^{2}}{2}g^{2}\left( 1-g^{2}\right) ^{2},  \label{13}
\end{equation}%
where $\lambda $ stands for the coupling constant for the scalar-matter
self-interaction (supposed to be dimensionless and positive). In this case,
the symmetric vacuum is $g=0$, the asymmetric one standing for $g=1$. Here,
we point out that the simplest self-dual Maxwell-Higgs model (saturated by
the usual $\left\vert \phi \right\vert ^{4}$-potential) presents only the
asymmetric vacuum $g=1$\ and does not support nontopological structures.
Within our generalized scenario, we circumvent this problem by adopting the $%
\left\vert \phi \right\vert ^{6}$-potential (\ref{13}), whilst choosing the
functions $G$\ and $w$\ conveniently.

Without loss of generality, and in the context of Lagrangian (\ref{1}), we
suppose\ that the function $G\left( g\right) $\ has the following behavior
at boundaries, i.e. when $g\left( r\right) \rightarrow 0$,%
\begin{equation}
G\left( g\right) =\frac{\gamma }{g^{2}}+\gamma _{0}+\gamma _{2}g^{2}...,  \label{Eq_G}
\end{equation}%
with the constants $\gamma $,$~\gamma _{0}$,$~\gamma _{2}$,$~...$\
characterizing such behavior. The case $\gamma \neq 0$\ allows to express
the total BPS energy as being proportional to the magnetic flux. In the
following we fix $\gamma =1$.

By using the Eqs. (\ref{13}) and (\ref{Eq_G}), we should investigate the way
the fields $g\left( r\right) $\ and $a\left( r\right) $ behave near the
boundaries. We thus solve the first-order equations (\ref{8}) and (\ref{9})
around the boundary values (\ref{3}) and (\ref{4}), obtaining the behavior
of profiles near the origin,
\begin{eqnarray}
g\left( r\right)  &\approx &C_{0}r^{n} , \label{g4} \\
a\left( r\right)  &\approx &n-\frac{\lambda C_{0}^{2}}{2\left( n+1\right) }%
r^{2n+2} , \label{g5}
\end{eqnarray}%
and asymptotically,
\begin{eqnarray}
g\left( r\right)  &\approx &\frac{C_{\infty }}{r^{\alpha _{n}}} ,\label{g6} \\
a\left( r\right)  &\approx &-\alpha _{n}+\frac{\lambda C_{\infty }^{2}}{%
2\left( \alpha _{n}-1\right) }\frac{1}{r^{2\alpha _{n}-2}}.
\label{g7}
\end{eqnarray}%
Here, $C_{0}$\ and $C_{\infty }$\ are positive real constants obtained
numerically by requiring proper behavior near and far from the origin,
respectively.

The point to be raised here is that the solutions (\ref{g4}) and (\ref{g5})
stand for a typical behavior of a topological vortex near the origin whereas
Eqs. (\ref{g6}) and (\ref{g7}) encode the typical nontopological profile far
from the origin. It is worthwhile to say that in Ref. \cite{cshv}, it was
shown that the very same dependence in $r$\ (despite some numerical factors)
characterizes the nontopological vortices of the usual Chern-Simons-Higgs
model.

For small $g$, replacing expressions (\ref{13}) and (\ref{Eq_G}) in the
self-dual Eqs. (\ref{8}) and (\ref{9}), one achieves a Liouville's equation
for $g\left( r\right) $ (in accordance with the procedure of Ref. \cite{cshv}%
), whose analytical solution is%
\begin{equation}
g\left( r\right) =\frac{2\left( n+1\right) }{r_{0}\sqrt{\lambda }}\frac{%
\left( \frac{r}{r_{0}}\right) ^{n}}{\left( \frac{r}{r_{0}}\right) ^{2\left(
n+1\right) }+1} , \label{ag}
\end{equation}%
with its maximum located at $r=r_{0}\left( n/\left( n+2\right) \right)
^{1/\left( 2n+2\right) }$\ (for $n$\ sufficiently large, its location
matches $r_{0}$). The corresponding expression for the gauge function $%
a\left( r\right) $\ reads as%
\begin{equation}
a\left( r\right) =-n-2+\frac{2\left( n+1\right) }{\left( \frac{r}{r_{0}}%
\right) ^{2\left( n+1\right) }+1}.  \label{aa}
\end{equation}%
The (approximate) solutions attained this way fulfill conditions (\ref{3})
and (\ref{4}), and hold for $r\rightarrow 0$\ or $r\rightarrow \infty $,
with $\alpha _{n}=n+2$, confirming the nontopological behavior.

\section{Some models endowed with nontopological solutions}

We now show\ some models belonging to the generalized Maxwell-Higgs
Lagrangian (\ref{1}), endowed with the sixth-order potential (\ref{13}).
These models are better characterized by choosing a specific function $%
w\left( g\right) $, from which the constraint (\ref{6}) yields $G\left(
g\right) $ behaving as (\ref{Eq_G}).

The first model is defined by the potential (\ref{13}) and the functions
\begin{equation}
w\left( g\right) =\frac{2}{3}\lambda \left( g^{2}+1\right)  ,\quad
G\left( g\right) =\frac{\left( g^{2}+3\right) ^{2}}{9g^{2}}.
\label{g1}
\end{equation}%
From Eqs. (\ref{8}) and (\ref{9}), we get the self-dual equations of this
model,%
\begin{equation}
\frac{dg}{dr}=\frac{ag}{r} ,\quad \frac{1}{r}\frac{da}{dr}=\frac{3\lambda g^{2}\left( g^{2}-1\right) }{g^{2}+3} , \label{g3}
\end{equation}
which was firstly studied in Ref. \cite{gv} (except for some numerical
factors) for analysis of topological solitons. The BPS nontopological
configurations of this specific model satisfy the boundary conditions (\ref%
{3}) and (\ref{4}), with energy given by Eq. (\ref{11a}). From the choices
of this model, we calculate $H\left( 0\right) =n\lambda $\ and $H\left(
\infty \right) =-\alpha _{n}\lambda $, so that the total energy is
\begin{equation}
E_{bps}=2\pi \lambda \left( n+\alpha _{n}\right) .  \label{ae}
\end{equation}%
thus we verify that the energy of the resulting self-dual structures is
proportional to their magnetic flux.

The second $|\phi |^{6}$-model we study is defined by the functions%
\begin{equation}
w(g)=\lambda  ,\quad G\left( g\right) =\frac{1}{g^{2}} ,\label{G}
\end{equation}%
whilst the self-dual equations are%
\begin{equation}
\frac{dg}{dr}=\frac{ag}{r} ,\quad \frac{1}{r}\frac{da}{dr}=\lambda
g^{2}\left( g^{2}-1\right)  , \label{e2}
\end{equation}
whose topological solutions were analyzed in Ref. \cite{gv}. We observe that
Eqs. (\ref{e2}) mimic those of the standard self-dual Chern-Simons-Higgs
model \cite{cshv} for $\lambda =2/\kappa ^{2}$\ ($\kappa $ standing for the
Chern-Simons constant). Despite satisfying the very same differential
equations, the solutions we describe in this work are physically different,
since they present no electric charge.

By solving the BPS equations (\ref{e2}) around the boundary values (\ref{3})
and (\ref{4}), we verify that $g\left( r\right) $\ and $a\left( r\right) $\
are given as in Eqs. (\ref{g4}) and (\ref{g5}) near the origin, while their
asymptotic behavior coincides with the ones of (\ref{g6}) and (\ref{g7}).
Thus, the solutions coming from (\ref{e2}) indeed stand for nontopological
self-dual vortices possessing no electric charge. The total BPS energy is
computed by using\ Eqs. (\ref{13}), (\ref{G}), (\ref{3}) and (\ref{4}),
leading to $H\left( 0\right) =n\lambda $, $H\left( \infty \right) =-\alpha
_{n}\lambda $, and $E_{bps}=2\pi \lambda \left( n+\alpha _{n}\right)
=\lambda \Phi _{B}$, being proportional to the\ magnetic flux again.

At this point, it is interesting to note that because of Eqs. (\ref{g3}) and
(\ref{e2}) the values of the winding number $n$\ are limited by the
requirement $g\leq 1$, otherwise we do not obtain solutions satisfying the
boundary conditions or with finite energy.

The third $|\phi |^{6}$-model to be scrutinized is one that affords
equations, which have an analytical solution. It is specified by the
functions
\begin{equation}
w=2\lambda \left( 1-g^{2}\right)  ,\quad G\left( g\right) =\frac{%
\left( 1-g^{2}\right) ^{2}}{g^{2}} , \label{w3}
\end{equation}%
providing the following self-dual equations:%
\begin{equation}
\frac{dg}{dr}=\frac{ag}{r} ,\quad \frac{1}{r}\frac{da}{dr}=-\lambda
g^{2}. \label{e2x}
\end{equation}
The Eqs. (\ref{e2x}) are exactly solvable, having analytical solutions given
by (\ref{ag}) and (\ref{aa}), which hold for all $r$,\ not only at origin or
at infinity. These solutions fulfill conditions (\ref{3}) and (\ref{4}),
with $\alpha _{n}=n+2$, possessing magnetic flux $\Phi _{B}=4\pi \left(
n+1\right) $ and total energy $E_{bps}=4\pi \lambda \left( n+1\right) $.

The requirement of a positive-definite energy density (\ref{BPS_def_pos}) or
positive $w\left( g\right) $ in (\ref{w3}) is equivalent to demanding $g\leq
1$, which holds only when%
\begin{equation}
\lambda r_{0}^{2}>n^{n/\left( n+1\right) }\left( n+2\right) ^{\left(
n+2\right) /\left( n+1\right) }  \label{ln}
\end{equation}%
is satisfied.\ Therefore, for a given value of the coupling constant $%
\lambda $, the possible values of the vorticity are limited in such a way
that larger values of $r_{0}$\ imply larger values of $n$.

We\ now show the numerical solutions for fixed $\lambda =100$ and $r_{0}=1$.
For these values, the winding number is restricted to $1\leq n\leq 4$, $%
1\leq n\leq 5,$\ or $1\leq n\leq 8,$\ for the first, the second and the
third models, respectively. It is worthwhile to note that a similar
situation holds in the Chern-Simons-Higgs model \cite{cshv}, whose the BPS
equations are the very same as those of the second model, see (\ref{e2}).
\begin{figure}
\includegraphics[width=8.6cm]{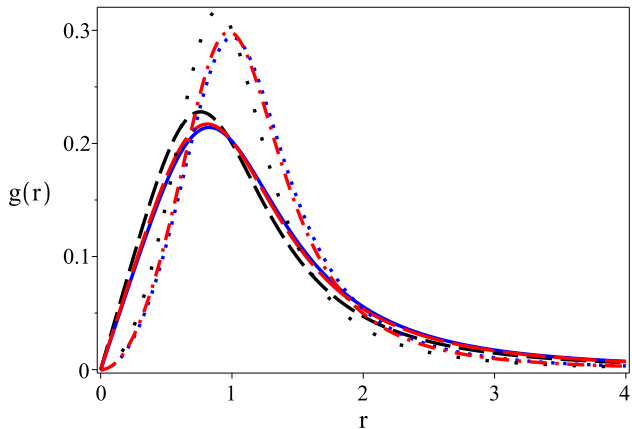}
\caption{Solutions to $g\left( r\right) $ calculated via (i) (\protect\ref%
{g3}) {\ (solid blue line for }${n=1}${, dotted blue line for }${n=2}${),
(ii) (\protect\ref{e2}) (long-dashed red line for }${n=1}${, dash-dotted red
line for }${n=2}${), and (iii) (\protect\ref{e2x}) (i.e., the analytical
profiles (\protect\ref{ag}) and (\protect\ref{aa}); dashed black line for }${%
n=1}${, space-dotted black line for }${n=2}$). We have used $\protect\lambda %
=100${\ and }$r_{0}=1${.}}
\end{figure}

In Figs. 1 and 2, we depict the solutions for $g\left( r\right) $ and $%
a\left( r\right) $, respectively. The first ones are peaked at rings around
the origin, whereas the second ones stand for peaks centered at the origin.
It is interesting to note that, for the first model one finds $\alpha
_{1}=3.03795$, $\alpha _{2}=4.07752$, and $\alpha _{1}=3.02977$, $\alpha
_{2}=4.06176$,\ for the second model. Therefore, we find a small deviation
of the value $\alpha _{n}=n+2$.
\begin{figure}
\includegraphics[width=8.6cm]{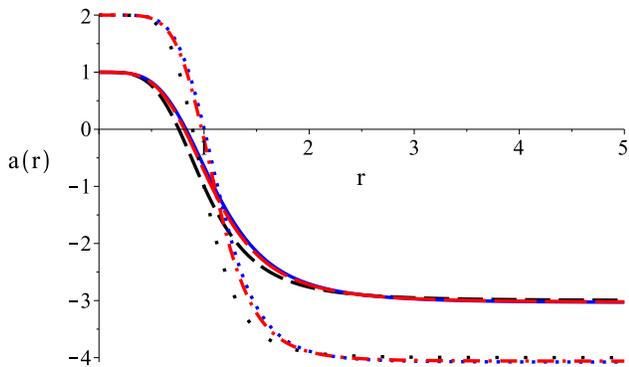}
\caption{Solutions to $a\left( r\right) $. Conventions as in Fig. 1.}
\end{figure}
\begin{figure}
\includegraphics[width=8.6cm]{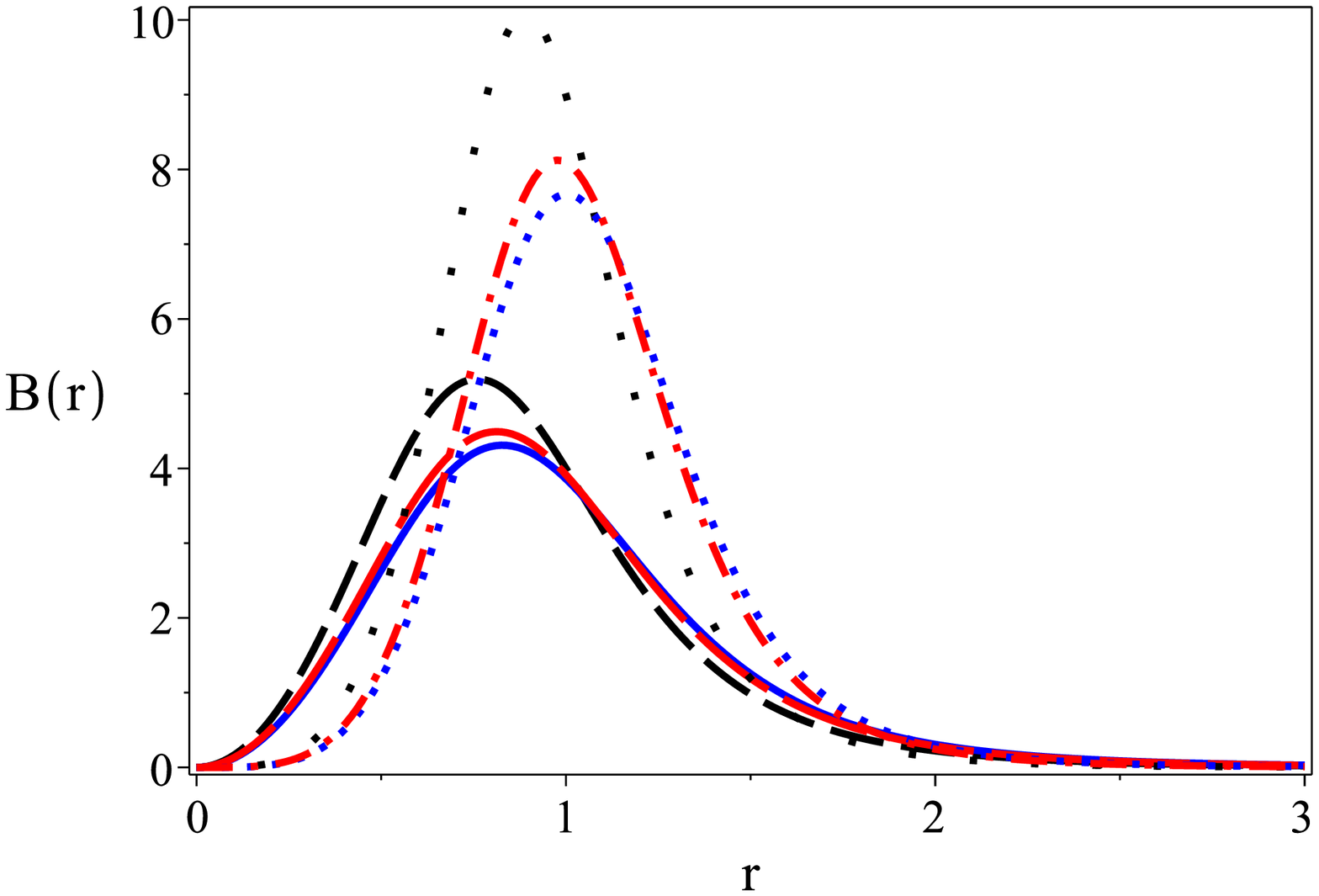}
\caption{Solutions to $B\left( r\right) $. Conventions as in Fig. 1.}
\end{figure}
\begin{figure}
\includegraphics[width=8.5cm]{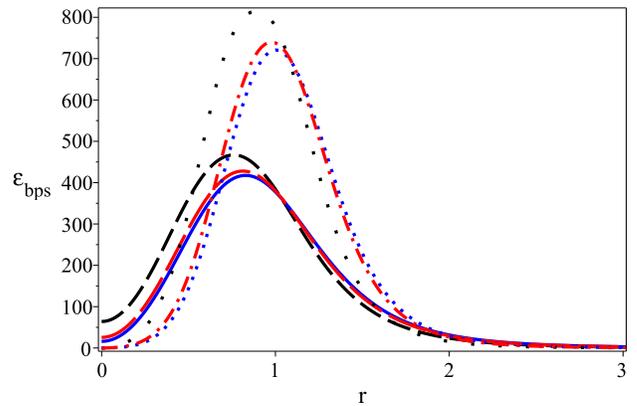}
\caption{Solutions to $\protect\varepsilon _{bps}$. Conventions as in Fig.
1. }
\end{figure}

The Figs. 3 and 4 show the solutions for the magnetic field $B\left(
r\right) =-(da/dr)/r$ and for the self-dual energy density (\ref{11}),
respectively. In general, all the profiles exhibit the same behavior, being
ring-like peaks centered at the origin, and vanishing asymptotically.


\section{Ending comments}

\label{end}

In this work, we have investigated the way the generalized Maxwell-Higgs
model (\ref{1}) gives rise to self-dual nontopological magnetic vortices
possessing finite energy and no electric charge. In order to present our
results, we have first reviewed the basic features of the model, which
engenders self-duality only when a particular constraint (\ref{6}) is
satisfied. We have specified our Maxwell-Higgs models by choosing the usual
sixth-order potential (\ref{13}) and changing the function $w\left(
\left\vert \phi \right\vert \right) $ conveniently, from which we have
obtained the corresponding expression for $G\left( \left\vert \phi
\right\vert \right) $ from the constraint (\ref{6}). Without loss of
generality, we have normalized the function $G(g)$ in such way that the
energy of the nontopological solutions is proportional to the corresponding
magnetic flux.

The BPS equations of the models here analyzed when $g\rightarrow 0$ reduce
to exactly the equations (\ref{e2x}). In this field regimen the solutions
are exact and given by Eqs. (\ref{ag}) and (\ref{aa}). These equations and
their solutions are invariant under the following scaling transformation,%
\begin{equation}
r\rightarrow \delta r\, ,~g\left( \delta r\right) =\delta ^{-1}g\left(
r\right)  ,~a\left( \delta r\right) =a\left( r\right)  ,
\end{equation}%
with $\delta $ the transformation parameter. Such an invariance enhances the
solutions of the models because there must exist a free parameter defining a
family of infinite solutions. In $\left\vert \phi \right\vert ^{6}$-models
defined in $\left( 1+2\right) $-dimensions the $\lambda $\ coupling constant
is dimensionless. Thus it becomes a free parameter characterizing the
solutions. For example, $\lambda $ determines the behavior of the solutions
of (\ref{e2x}) given by Eqs. (\ref{ag}) and (\ref{aa}) when $g\rightarrow 0$%
. This way, when $r\rightarrow 0$: 
\begin{equation}
g\left( r\right) \simeq \frac{2\left(
n+1\right) }{r_{0}\sqrt{\lambda }}\left( \frac{r}{r_{0}}\right) ^{n}+...,
\end{equation}
and for $r\rightarrow \infty $: 
\begin{equation}
g\left( r\right) \simeq \frac{2\left(
n+1\right) }{r_{0}\sqrt{\lambda }}\left( \frac{r_{0}}{r}\right) ^{n+2}+....
\end{equation}

An important result is that for fixed  $\lambda $ and $r_{0}$, we can
determine the values of $n~$\ for which exist nontopological BPS vortices,
as given by Eq. (\ref{ln}). In fact, we have verified that the winding
number values are limited by the condition $g\leq 1$. Despite this
inequality (\ref{ln}) is exact only for the solution of the set of BPS
equations (\ref{e2x}) and it gives an upper-bound for the values of $n$  in
all models we have considered.

An other result is related to the BPS energy of the nontopological vortices;
for the models we have studied, the energy depends explicitly on the free
parameter $\lambda $, see (\ref{ae}).

The stability issue of the new nontopological BPS solutions is under
investigation and the results will be reported elsewhere.

An other interesting issue concerns the search for BPS solutions in
generalized Born-Infeld-Higgs models. This issue is now under consideration,
and we expect interesting results for a future contribution.

\acknowledgments The authors thank CAPES, CNPq and FAPEMA (Brazilian
agencies) for partial financial support.

\end{document}